\documentclass[12pt]{article}
 \usepackage{graphicx,amsfonts,amsmath}

 \renewcommand{\title}[1] {%
 \begingroup\begin{center}\vspace{0.0cm}\bf\Large
 \addtolength{\baselineskip}{1mm} #1 \end{center}\endgroup}

 \renewcommand{\author}[1] {%
 \begingroup\begin{center}\vspace{0.2cm}\bf #1 \vspace{0.2cm}
 \end{center}\endgroup}

  \newcommand{\address}[1] {%
 \begingroup\begin{center} #1 \end{center}\endgroup}

 \newcommand\be{\begin{equation}}
\newcommand\ee{\end{equation}}
\newcommand\bee{\begin{eqnarray}}
\newcommand\eee{\end{eqnarray}}
\newcommand{\e}{\mbox{e}}
\newcommand{\Pf}{\mbox{Pf}\,}

\begin{document}

 \title{Correlation function of the two-dimensional Ising model
 on a finite lattice. II}
 \author{A. I. Bugrij$^{\; *}$, O. Lisovyy$^{\;*,\;\dag}$}
 \address{
  $^{*\;}$Bogolyubov Institute for Theoretical Physics \\
  Metrolohichna str., 14-b, Kyiv-143, 03143, Ukraine \vspace{0.2cm} \\
  $^{\dag\;}$ Laboratoire de Math\'ematiques et Physique Th\'eorique CNRS/UMR 6083,\\
  Universit\'e de Tours, Parc de Grandmont, 37200 Tours, France}
  \date{}

 \begin{abstract}
 We calculate the two-point correlation function and magnetic
 susceptibility in the anisotropic 2D Ising model on a lattice
 with one infinite and the other finite dimension, along which
 periodic boundary conditions are imposed.
 Using exact expressions for a part of lattice form factors, we propose
 the formulas for arbitrary spin matrix elements, thus providing a possibility
 to compute all multipoint correlation functions in the anisotropic Ising model
 on cylindrical and toroidal lattices. The scaling limit of the
 corresponding expressions is also analyzed.
 \end{abstract}
 \section{The model}
 Two-point correlation function and magnetic susceptibility of the isotropic
 Ising model on a cylinder were calculated in \cite{bugrij1, we}.
 Analogous results can be obtained in the anisotropic case as
 well. The complications due to the presence of two
 different coupling constants can be overcome, since the matrix
 whose determinant yields the correlation function continues to
 have a Toeplitz form. The computation idea has already been presented
 in \cite{bugrij1}, and below we will often use the results of that paper,
 omitting detailed calculations.

 The calculation of multipoint correlation functions can be
 reduced to the problem of finding Ising spin matrix elements (the
 so-called form factors) in
 the orthonormal basis of transfer matrix eigenstates. Although
 neither the first nor the second problem has been solved on a
 finite lattice, the language of matrix elements turns out to be
 more convenient when constructing the corresponding expressions.
 A part of form factors can be found from the formulas
 for the two-point correlation function. Since the structure of
 these expressions is relatively simple, we generalize them to the
 case of arbitrary matrix elements.

 The hamiltonian of the anisotropic Ising model on a rectangular lattice
 is defined as
 $$H[\sigma]=-\sum\limits_{\mathbf{r}}\sigma(\mathbf{r})
 \left(J_x \nabla_x+J_y \nabla_y\right)\sigma(\mathbf{r}),$$
 where the two-dimensional vector $\mathbf{r}=(r_x,r_y)$ labels
 the lattice sites: $r_x=1,\;2,\;\ldots,$ $\;M,$ $r_y=1,\;2,\;\ldots,\;N$;
 the spins $\sigma(\mathbf{r})$ take on the values $\pm1$.
 The parameters $J_x$ and $J_y$ determine the coupling energies of adjacent spins
 in the horizontal and vertical direction. The operators of shifts
 by one lattice site,
 $\nabla_x$ and $\nabla_y$, are given by
 $$\nabla_x \sigma(r_x,r_y)=\sigma(r_x+1,r_y),\;\;\;\nabla_y \sigma(r_x,r_y)=\sigma(r_x,r_y+1),$$
 where for periodic boundary conditions one has
 $$\left(\nabla_{x}^{(R)}\right)^{M}=\left(\nabla_{y}^{(R)}\right)^{N}=1,$$
 and for antiperiodic ones
 $$\left(\nabla_{x}^{(NS)}\right)^{M}=\left(\nabla_{y}^{(NS)}\right)^{N}=-1.$$

 If the lattice is periodic in both directions, the partition function
 of the model at the temperature $\beta^{-1}$
 $$Z=\sum_{[\sigma]}e^{-\beta H[\sigma]}$$
 can be written as a sum of four terms
 \be\label{0} Z=
 \frac12 \left(Q^{(NS,\,NS)}+Q^{(NS,\,R)}+Q^{(R,\,NS)}
 -Q^{(R,\,R)}\right),\ee
 each of them being proportional to the pfaffian of the operator
 $$\hat{D}=\left(\begin{array}{cccc}
   0 & 1 & {1+t_x \nabla_x} & 1 \\
   -1 & 0 & 1 & {1+t_y \nabla_y} \\
   {-1-t_x \nabla_{-x}} & -1 & 0 & 1 \\
   -1 & {-1-t_y \nabla_{-y}} & -1 & 0 \ \end{array}\right)\,, $$
   $$Q=\left(2\cosh K_x \cosh K_y\right)^{MN}\cdot\mbox{Pf}\,\hat{D}\,,$$
 with different boundary conditions for $\nabla_x$, $\nabla_y$.
 Here, we have introduced the dimensionless parameters
 $$K_x=\beta J_x,\;\;\;K_y=\beta J_y,\;\;\;t_x=\tanh K_x,\;\;\;t_y=\tanh K_y.$$
 One can verify that when the torus degenerates into a cylinder ($M\gg N$), the partition
 function is determined by the antiperiodic term:
 $Z=Q^{(NS,\,NS)}$.

 Let us consider the two-point correlation function in the Ising model on the
 cylinder
 \be\langle\sigma(0,0)\sigma(r_x,r_y)\rangle=Z^{-1}\sum_{[\sigma]}
 \sigma(0,0)\sigma(r_x,r_y)
 \e^{-\beta
 H[\sigma]}\label{1}{}=i^{P}\frac{Z_{def}}{Z}
 =i^{P}\frac{Q^{(NS,\,NS)}_{def}}{Q^{(NS,\,NS)}},\ee
 where $Z_{def}$ denotes the partition function of the Ising model with a defect:
 the coupling parameters $K_{x}$, $K_{y}$ should be replaced by $K_{x}-i\pi/2$,
 $K_{y}-i\pi/2$ along a path that connects the correlating spins (see Fig.~1 in \cite{bugrij1}, where the
 numbering of lattice sites and the locations of correlating spins were described; the bold line in the
 figure is the path along which the couplings are modified).
 The exponent $P$ in the formula (\ref{1}) is equal to the number of steps along the
 defect line:
 $P=r_{x}+r_{y}$ in the case of a shortest path.
 When the correlating spins are located along a line parallel to the cylinder axis (i.~e. $r_y=0$),
 the ratio of pfaffians in the right hand side of (\ref{1})
 can be expressed in terms of the determinant of a Toeplitz
 matrix
 \be\label{2} \langle\sigma(0,0)\sigma(r_x,0)\rangle =\mathrm{det}\,A,\ee
\be\label{3}
 A_{kk'}=\int\limits_{-\pi}^{\pi}\frac{dp\;e^{ip(k'-k)}}{2\pi N}\;{\sum_{q}}^{(NS)}
 \frac{2t_x\left(1+t_y^2\right)+\left(t_y^2-1\right)\left(e^{-ip}+t_x^2e^{ip}\right)}
 {\left(1+t_x^2\right)\left(1+t_y^2\right)-2t_x\left(1-t_y^2\right)\cos p-
 2t_y\left(1-t_x^2\right)\cos q},\ee
 whose size $|r_x|\times|r_x|$ is determined by the distance between the correlating spins.
 Here and below the superscripts $(NS)$ and $(R)$ in sums and products imply that the corresponding
 operations are performed with respect to Neveu-Schwartz
 ($q=\frac{2\pi}{N}\,\left(j+\frac12\right)$, $j=0,1,\ldots,N-1$)
 or Ramond ($q=\frac{2\pi}{N}\,j$, $j=0,1,\ldots,N-1$)
 values of quasimomenta.
 Our goal is to transform (\ref{2})--(\ref{3}) into a representation with an explicit
 dependence on the distance.

 \section{Ferromagnetic phase}
 In the translationally invariant case the pfaffians $\Pf \hat{D}$
 can be easily calculated,
 \be\label{4a} Q=\prod_{q,p}(\cosh 2K_{x}\cosh 2K_{y}-
 \sinh 2K_{x}\cos q-\sinh 2K_{y}\cos
 p)^{1/2}.\ee
 The product over any of the two quasimomentum components in
 (\ref{4a}) can be found in an explicit form. For instance, the term in
 (\ref{0}),
 which corresponds to periodic boundary conditions along the $x$ axis and antiperiodic ones
 along the $y$ axis, may be written as
  \begin{eqnarray}
 \nonumber Q^{(R,NS)}&=&(2\sinh2K_{y})^{MN/2}{\prod_{q}}^{(R)}\e^{N\gamma(q)/2}
 (1+\e^{-N\gamma(q)})=\\
 \label{4b}{}&=&(2\sinh2K_{x})^{MN/2}{\prod_{q}}^{(NS)}\e^{M\overline{\gamma}(p)/2}
 (1-\e^{-M\overline{\gamma}(p)}),\end{eqnarray}
 where the functions $\gamma(q)$ and $\overline{\gamma}(p)$ are determined by the relations
 \be\label{4}\cosh\gamma(q)=\frac{(1+t_x^2)(1+t_y^2)}{2t_x (1-t_y^2)}-
 \frac{t_y(1-t_x^2)}{t_x (1-t_y^2)}\,\cos q,\ee
 \be\label{5}\cosh\overline{\gamma}(p)=\frac{(1+t_x^2)(1+t_y^2)}{2t_y (1-t_x^2)}-
 \frac{t_x(1-t_y^2)}{t_y (1-t_x^2)}\,\cos p\ee
 and the conditions $\gamma(q),\overline{\gamma}(p)>0$.

 We will call the domain of values of  $K_x$, $K_y$,
 where
 $$\sinh 2K_x\;\sinh 2K_y>1$$
 (and hence $\gamma(0)=\ln t_x+2K_y$) the ferromagnetic region of parameters. Notice that in this
 case the numerator of the integrand in (\ref{3}) can be represented in the following
 factorized form in terms of the function $\gamma(q)$:
 $$2t_x\left(1+t_y^2\right)+\left(t_y^2-1\right)\left(e^{-ip}+t_x^2e^{ip}\right)
 =(1-t_y^2)e^{\gamma(0)}
 (1-e^{-\gamma(\pi)}e^{ip})
 (1-e^{-\gamma(0)}e^{-ip}).$$
 Then,  using the identity
 \be\label{6}\frac1N\;{\sum_{q}}^{(NS)}\frac{1}{\cosh \theta-\cos q}=\frac{\tanh\frac{N\theta}{2}}{\sinh\theta},
 \ee
 one may compute the sum over the discrete Neveu-Schwartz spectrum in (\ref{3}).
 As a result, we can express matrix elements
 $A_{kk'}$ in terms of contour integrals
 \be\label{7}A_{kk'}=\frac{1}{2\pi i}\oint\limits_{|z|=1}\frac{dz}{z}\,z^{k'-k}A(z),\ee
 \be\label{8}A(z)=\left[\frac{(1-e^{-\gamma(\pi)}z)(1-e^{-\gamma(0)}z^{-1})}
 {(1-e^{-\gamma(0)}z)(1-e^{-\gamma(\pi)}z^{-1})}\right]^{\frac12}\,T(z),\ee
 where
 \be\label{9}T(z)=\tanh\frac{N\overline{\gamma}(p)}{2},\;\;\;z=e^{ip}.\ee

 Recall that in order to calculate the determinant $|A|$ by the
 Wiener-Hopf method
 \cite{McCoyWu} one has to represent the kernel $A(z)$
 in a factorized form
 \be\label{10}A(z)=P(z)Q(z^{-1}),\ee
 where the functions $P(z)$ and $Q(z)$ are analytic inside the unit circle.
 Setting
 \be\label{11}W(z)=\frac{P(z)}{Q(z^{-1})}\ee
 the determinant can be written as follows:
 \be\label{12}|A|=e^{h(r_x)}\sum\limits_{l=0}^{\infty}g_{2l}(r_x),\ee
 where
 \be\label{13}h(r_x)=r_x\left[\ln P(0)+\ln Q(0)\right]+\frac{1}{2\pi i}\oint\limits_{|z|=1}dz
 \ln Q(z^{-1})\frac{\partial}{\partial z}\ln P(z),\ee
 \be\label{14}g_{2l}(r_x)=\frac{(-1)^l}{l!l!(2\pi i)^l}\oint\limits_{|z_i|<1}
 \frac{\prod\limits_{i=1}^{2l}(dz_i\,z_i^{r_x})\prod\limits_{i=1}^{l-1}\prod\limits_{j=i+1}^{l}
 \left[(z_{2i-1}-z_{2j-1})^2(z_{2i}-z_{2j})^2\right]}
 {\prod\limits_{i=1}^{l}\prod\limits_{j=1}^{l}(1-z_{2i-1}z_{2j})^2}\,
 \frac{\prod\limits_{i=1}^{l}W(z_{2i-1})}{\prod\limits_{i=1}^{l}W(z_{2i}^{-1})}\,.\ee
 We will see that the sum (\ref{12}) contains a finite number of
 terms, i.~e. there exists $l_0$ such that $g_{2l}\equiv
 0$ for all $l>l_0$.

 Let us now calculate the integral (\ref{14}). To do this, we first rewrite
 the function $T(z)$ defined by (\ref{9}) in the factorized form:
 $$T(z)=\tanh\frac{N\overline{\gamma}(p)}{2}=\left[
 \frac{{\prod\limits_{q}}^{(R)}(\cosh\overline{\gamma}(p)-\cos q)}
 {{\prod\limits_{q}}^{(NS)}(\cosh\overline{\gamma}(p)-\cos
 q)}\right]^{\frac12}=$$
  $$=\left[ \frac{{\prod\limits_{q}}^{(R)}
 e^{\gamma(q)}(1-e^{-\gamma(q)}z)(1-e^{-\gamma(q)}z^{-1})}
 {{\prod\limits_{q}}^{(NS)}
 e^{\gamma(q)}(1-e^{-\gamma(q)}z)(1-e^{-\gamma(q)}z^{-1})}
 \right]^{\frac12}=$$
 $$=P_T(z)Q_T(z^{-1})=e^{-\Lambda^{-1}}Q_T(z)\,Q_T(z^{-1}).$$
 Here, we have introduced the notation
 $$\Lambda^{-1}=\frac12\left[{\sum\limits_q}^{(NS)}\gamma(q)-{\sum\limits_q}^{(R)}\gamma(q)\right],
 $$
 \be\label{15}Q_T(z)=\left[ \frac{{\prod\limits_{q}}^{(R)}(1-e^{-\gamma(q)}z)}
 {{\prod\limits_{q}}^{(NS)} (1-e^{-\gamma(q)}z)} \right]^{\frac12},\;\;\;
 P_T(z)=e^{-\Lambda^{-1}}\;Q_T(z).\ee
 Next let us define
 $$P_0(z)=\left[\frac{1-e^{-\gamma(\pi)}z}{1-e^{-\gamma(0)}z}\right]^{\frac12},\;\;\;
 Q_0(z)=\left[\frac{1-e^{-\gamma(0)}z}{1-e^{-\gamma(\pi)}z}\right]^{\frac12},$$
 and write $W(z)$ as follows:
 \begin{eqnarray*}
 W(z)&=&\frac{P_0(z)}{Q_0(z^{-1})}\,\frac{P_T(z)}{Q_T(z^{-1})}=\frac{P_0(z)}{Q_0(z^{-1})}
 \,\frac{Q_T^2(z)}{T(z)}=\\
 &=&\frac{t_x(1-t_y^2)}{t_y(1-t_x^2)}\,
 (\cosh\gamma(\pi)-\cos p)\,Q_T^2(z)\,\frac{\coth\frac{N\overline{\gamma}(p)}{2}}
 {\sinh\overline{\gamma}(p)}\;e^{\frac{\gamma(0)-\gamma(\pi)}{2}\,-\,2\Lambda^{-1}}\;,
 \end{eqnarray*}
 where $z=e^{ip}$. Taking into account that
 \be\label{16}\frac{\coth\frac{N\overline{\gamma}(p)}{2}}{\sinh\overline{\gamma}(p)}=
 \frac1N\,\frac{t_y(1-t_x^2)}{t_x(1-t_y^2)}\;{\sum_{q}}^{(R)}\frac{1}{\cosh \gamma(q)-\cos p}\,,
 \ee
  one can write $W(z)$ as
 $$W(z)=(\cosh\gamma(\pi)-\cos p)\,Q_T^2(z)\;
 e^{\frac{\gamma(0)-\gamma(\pi)}{2}\,-\,2\Lambda^{-1}}\;
 \frac1N\;{\sum_{q}}^{(R)}\frac{1}{\cosh \gamma(q)-\cos p}\,.$$
 Analogous computation for $W^{-1}(z^{-1})$ gives
 $$W^{-1}(z^{-1})=(\cosh\gamma(0)-\cos p)\,Q_T^2(z)\;
 e^{\frac{\gamma(\pi)-\gamma(0)}{2}}\;
 \frac1N\;{\sum_{q}}^{(R)}\frac{1}{\cosh \gamma(q)-\cos p}\,.$$
 Using the last two formulas, we can calculate the integral (\ref{14}).
 Let us interchange the order of integration and summation over discrete quasimomenta.
 Then all the singularities inside the integration contours are exhausted by a finite
 number of poles at the points
 $$z_{(k)}=e^{-\gamma(q_{(k)})},\;\;\;q_{(k)}=\frac{2\pi k}{N}\,,\;\;\;
 k=0,\;1,\;\ldots,\;N-1,$$
 corresponding to the Ramond values of quasimomentum. The integral (\ref{14})
 can then be computed by residues, and the result reads
 $$g_{2l}(r_x)=\frac{e^{-2l/\Lambda}}{(2l)!N^{2l}}\;{\sum\limits_q}^{(R)}
 \prod\limits_{i=1}^{2l}\frac{e^{-r_x
 \gamma(q_i)-\eta(q_i)}}{\sinh\gamma(q_i)}\;
 \mathcal{G}_{2l}[q],$$
 where
 \be\label{17}e^{-\eta(q_i)}=\frac{{\prod\limits_q}^{(R)}
 \left(1-e^{-\gamma(q)-\gamma(q_{i})}\right)}{{\prod\limits_q}^{(NS)}
 \left(1-e^{-\gamma(q)-\gamma(q_{i})}\right)}\;.\ee
 The function $\mathcal{G}_{2l}[q]$ is given by
 \begin{eqnarray*}\mathcal{G}_{2l}[q]&=&C_l^{2l}\prod\limits_{i=1}^{l-1}\prod\limits_{j=i+1}^{l}
 \left[\left(e^{-\gamma(q_{2i-1})}-e^{-\gamma(q_{2j-1})}\right)^2
 \left(e^{-\gamma(q_{2i})}-e^{-\gamma(q_{2j})}\right)^2\right]\;\times\\
 &\;&\times\;\frac{\prod\limits_{i=1}^{l}\left[
 \biggl(\cosh\gamma(\pi)-\cosh\gamma(q_{2i-1})\biggr)
 \biggl(\cosh\gamma(q_{2i})-\cosh\gamma(0)\biggr)\right]}
 {\prod\limits_{i=1}^{2l}e^{\gamma(q_i)}\prod\limits_{i=1}^{l}\prod\limits_{j=1}^{l}
 \left(1-e^{-\gamma(q_{2i-1})-\gamma(q_{2j})}\right)^2}=\\
 &=&\frac{C_l^{2l}}{2^{2l^2}}\;\left[\frac{t_y\,(1-t_x^2)}{t_x(1-t_y^2)}\right]^{2l^2}
 \;\frac{\prod\limits_{i=1}^{l-1}\prod\limits_{j=i+1}^{l}
 \left[(\cos q_{2i-1}-\cos q_{2j-1})^2(\cos q_{2i}-\cos q_{2j})^2\right]}
 {\prod\limits_{i<j}^{2l}\sinh^2\frac{\gamma(q_i)+\gamma(q_j)}{2}}\times\\
 &\;&\times\; \prod\limits_{i=1}^{l}\left[(1+\cos q_{2i-1})(1-\cos q_{2i})\right].
 \end{eqnarray*}
 It can be symmetrized with respect to the permutations $q_{2i}\leftrightarrow q_{2j-1}$.
 It turns out that the result of such a symmetrization coincides with the even
 part of the function
 $$ \left[\frac{t_y\,(1-t_x^2)}{t_x(1-t_y^2)}\right]^{2l^2} \prod\limits_{i<j}^{2l}
 \frac{\sin^{2}\frac{q_i-q_j}{2}}{\sinh^2\frac{\gamma(q_i)+\gamma(q_j)}{2}},$$
 so that $g_{2l}(r_x)$ may be written as
 \be\label{18}g_{2l}(r_x)=\frac{e^{-2l/\Lambda}}{(2l)!N^{2l}}
 \left[\frac{t_y\,(1-t_x^2)}{t_x(1-t_y^2)}\right]^{2l^2}\;
 {\sum\limits_{q}}^{(R)}\frac{e^{-|r_x|\gamma(q_i)-\eta(q_i)}}{\sinh\gamma(q_i)}\;
 F_{2l}^2[q],\;\;\;g_0=1,\ee
 \be\label{19}F_{n}[q]=\prod\limits_{i<j}^{n}
 \frac{\sin\frac{q_i-q_j}{2}}{\sinh\frac{\gamma(q_i)+\gamma(q_j)}{2}}\;.\ee
 The formula (\ref{19}) implies that for $2l>N$ $F_{2l}[q]\equiv0$.
 By virtue of the relation
 (\ref{12}), we then obtain the following representation for the correlation
 function in the ferromagnetic region:
 \be\label{20}\langle\sigma(0,0)\sigma(r_x,0)\rangle^{(-)}=\xi\xi_T\;e^{-|r_x|/\Lambda}\;
 \sum\limits_{n=0}^{[N/2]}g_{2n}(r_x),\ee
 where
 \be\label{21}\xi=\left|1-(\sinh2K_x\sinh2K_y)^{-2}\right|^\frac14, \ee
 \be\label{22}\xi_T=\left[\frac
 {{\prod\limits_{q}}^{(R)}{\prod\limits_{p}}^{(NS)}\sinh^2 \frac{\gamma(q)+\gamma(p)}{2}}
 {{\prod\limits_{q}}^{(R)}{\prod\limits_{p}}^{(R)}\sinh
 \frac{\gamma(q)+\gamma(p)}{2}\;
 {\prod\limits_{q}}^{(NS)}{\prod\limits_{p}}^{(NS)}\sinh \frac{\gamma(q)+\gamma(p)}{2}}
 \right]^{\frac14}.\ee

 \section{Paramagnetic phase}
 In the paramagnetic region of parameters, determined by the condition
  $\sinh2K_x\,\sinh2K_y<1$, we have the inequality
 $\ln t_x+2K_y<0$, and hence
 $$\gamma(0)=-\ln t_x-2K_y\,.$$
 Because of this, the kernel of the Toeplitz matrix whose determinant
 yields the correlation function should be rearranged:
 $$\langle\sigma(0,0)\sigma(r_x,0)\rangle^{(+)}=|A|,$$
 \be\label{23}A_{kk'}=\frac{1}{2\pi
 i}\oint\limits_{|z|=1}\frac{dz}{z}\,z^{k'-k-1}A(z),\ee
 \be\label{24}A(z)=\left[\frac{(1-e^{-\gamma(\pi)}z)(1-e^{-\gamma(0)}z)}
 {(1-e^{-\gamma(0)}z^{-1})(1-e^{-\gamma(\pi)}z^{-1})}\right]^{\frac12}\,T(z).
 \ee
 To calculate the determinant of such a matrix, we must modify
 the method described in the previous section. For the factorized kernel
 $$A(z)=P(z)Q(z^{-1}),$$
 where the functions $\ln P(z)$ and $\ln Q(z)$ are analytic inside the unit circle $|z|<1$,
 we have the following general formula \cite{bugrij1}
 \be\label{25}|A|=\frac{(-1)^{r_x}e^{h(r_x+1)}}{Q^2(0)}\;\sum\limits_{l=0}^{\infty}f_{2l+1}(r_x),
 \ee
 $$f_{2l+1}(r_x)=\frac{(-1)^l}{l!(l+1)!(2\pi i)^{2l+1}}\;\oint\limits_{|z_i|<1}
 \prod\limits_{i=1}^{2l+1}\left(dz_i\,z_i^{r_x}\right)\;
 \frac{\prod\limits_{i=1}^{l}\biggl(z_{2i}W(z_{2i})\biggr)}
 {\prod\limits_{i=0}^{l}\biggl(z_{2i+1}W(z_{2i+1}^{-1})\biggr)}\times$$
 \be\label{26}\times\;\frac{\prod\limits_{i=1}^{l-1}\prod\limits_{j=i+1}^{l}(z_{2i}-z_{2j})^2
 \prod\limits_{i=1}^{l}\prod\limits_{j=i+1}^{l+1}(z_{2i-1}-z_{2j-1})^2}
 {\prod\limits_{i=1}^{l}\prod\limits_{j=0}^{l}(1-z_{2i}z_{2j+1})^2}\;.\ee
 Here, $W(z)$ and $h(r_x)$ are defined by the formulas (\ref{11}) and (\ref{13}).
 Using the transformations analogous to those described above, one finds
 $$zW(z)=2z\,e^{-\frac{\gamma(0)+\gamma(\pi)}{2}\,-\,2/\Lambda}Q_{T}^2(z)
 (\cosh\gamma(0)-\cos p)(\cosh\gamma(\pi)-\cos p)\,\times$$ $$\times\;\frac1N\;{\sum\limits_{q}}^{(R)}
 \frac{1}{\cosh\gamma(q)-\cos p}\;,$$
 $$z^{-1}W^{-1}(z^{-1})=\frac{z^{-1}}{2}\;e^{\frac{\gamma(0)+\gamma(\pi)}{2}}\,Q_{T}^2(z)
 \,\frac1N\;{\sum\limits_{q}}^{(R)} \frac{1}{\cosh\gamma(q)-\cos p}\;.$$
 Using the last two relations, the integral (\ref{26})
 can be computed by residues:
 $$f_{2l+1}=\left[\frac{t_y(1-t_x^2)}{t_x(1-t_y^2)}\right]^{2l^2+2l}
 \frac{e^{-2l/\Lambda+\,\frac{\gamma(0)+\gamma(\pi)}{2}}}{(2l+1)!N^{2l+1}}\;
 {\sum\limits_q}^{(R)}\prod\limits_{i=1}^{2l+1}
 \frac{e^{-r_x\gamma(q_i)-\eta(q_i)}}{\sinh\gamma(q_i)}\,
 \frac{\mathcal{F}[q]}{\prod\limits_{i<j}^{2l+1}\sinh^2\frac{\gamma(q_i)+\gamma(q_j)}{2}}\;,$$
 where
 $$\mathcal{F}[q]=\frac{C^{2l+1}_l}{2^{2l(l+1)}}\;
 \prod\limits_{i=1}^{l-1}\prod\limits_{j=i+1}^{l}(\cos q_{2i}-\cos q_{2j})^2
 \prod\limits_{i=1}^{l}\prod\limits_{j=i+1}^{l+1}(\cos q_{2i-1}-\cos q_{2j-1})^2
 \prod\limits_{i=1}^{l}\sin^2 q_{2i}\;.$$
 We again note that after symmetrization over the permutations
 $q_{2i}\leftrightarrow q_{2j+1}$, the function $\mathcal{F}[q]$
 coincides with the even part of the function
 $$\prod\limits_{i<j}^{2l+1}\sin^2\frac{q_i-q_j}{2}\;.$$
 The correlation function in the paramagnetic region of parameters is then given by
 \be\label{27}\langle\sigma(0,0)\sigma(r_x,0)\rangle^{(+)}=\xi\xi_T\;
 e^{-|r_x|/\Lambda}\;
 \sum\limits_{n=0}^{[(N-1)/2]}g_{2n+1}(r_x),\ee
 \be\label{28}g_{2l+1}(r_x)=\frac{e^{-(2l+1)/\Lambda}}{(2l+1)!N^{2l+1}}
 \left[\frac{t_y\,(1-t_x^2)}{t_x(1-t_y^2)}\right]^{(2l+1)^2/2}\;
 {\sum\limits_{q}}^{(R)}\frac{e^{-|r_x|\gamma(q_i)-\eta(q_i)}}{\sinh\gamma(q_i)}\;
 F_{2l+1}^2[q],\ee
 \be\label{29}F_{n}[q]=\prod\limits_{i<j}^{n}
 \frac{\sin\frac{q_i-q_j}{2}}{\sinh\frac{\gamma(q_i)+\gamma(q_j)}{2}}\;,\;\;\;F_1=1.\ee

 Comparing the formulas (\ref{18})--(\ref{22}) and
 (\ref{27})--(\ref{29}) for the correlation function with the
 isotropic case, we see that all the difference consists in the
 redefinition of the function $\gamma(q)$ given by
 (\ref{4}) and in the appearance of the multipliers
 $$\frac{t_y(1-t_x^2)}{t_x(1-t_y^2)}$$
 in the appropriate powers in the function
  $g_{n}(r_x)$. In the thermodynamic limit $N\rightarrow\infty$, these expressions
  reduce to the known formulas  \cite{yamada}.

 \section{Magnetic susceptibility}
 Since we have two independent coupling parameters in the
 anisotropic Ising model on the cylinder, we can consider two limit
 cases:
 $$K_{x}\to0, \qquad K_{y}\neq0$$
  and
  $$K_{x}\neq0,\qquad K_{y}\to0.$$
  The two-dimensional Ising lattice splits into mutually non-interacting closed one-row
  Ising chains of the length $N$ in the first case, and into $N$ chains of
  infinite length in the second case. Such limit cases provide primitive tests
  for the form factor representations obtained above. To perform such checks, it would also
  be desirable
  to have an expression for the correlation function $\langle\sigma(0)\sigma(\bf{r})\rangle$
  of two spins with arbitrary location on the lattice, i.~e. for $r_x\neq0$
  and $r_y\neq0$. The same problem arises when one tries to write the lattice analog of the
  Lehmann representation for the two-point Green function or to compute the magnetic susceptibility:
  in both cases we have to perform the Fourier transformation
$\sum_{\bf{q}}\e^{i\bf{q}\bf{r}}\langle\sigma(0)\sigma(\bf{r})\rangle$.
 Meanwhile, the matrix whose determinant gives the correlation function
 has a non-Toeplitz form for $r_y\neq 0$, and the method used above to find form factor expansions
 becomes inapplicable. The answer can nevertheless be obtained in this case as well
 if we formulate the problem in terms of spin matrix elements.

 Recall that the $2^N$-dimensional space in which the Ising transfer matrix
 $\mathcal{T}$ acts can be splitted into two invariant subspaces  --- the
 Neveu-Schwartz sector (NS) and the Ramond sector (R). The
 eigenvalues corresponding to the eigenvectors from each subspace
 are given by
  \begin{eqnarray}\label{30}
 \lambda^{(NS)}&=&(2\sinh K_x)^{N/2}\exp\frac12\left\{\pm\gamma\left(\frac{\pi}{N}\right)
 \pm\gamma\left(\frac{3\pi}{N}\right)\pm\ldots
 \pm\gamma\left(2\pi-\frac{\pi}{N}\right)\right\},\\
 \label{31}
 \lambda^{(R)}&=&(2\sinh K_x)^{N/2}\exp\frac12\left\{\pm\gamma\left(0\right)
 \pm\gamma\left(\frac{2\pi}{N}\right)\pm\ldots
 \pm\gamma\left(2\pi-\frac{2\pi}{N}\right)\right\},\end{eqnarray}
 and we have different selection rules in different regions of parameters of the
 model: the number of minus signs in
 the NS--sector (\ref{30}) is even in both ferromagnetic and paramagnetic region,
 while the number of minus signs in
 the R--sector (\ref{31})
 is even in the ferromagnetic region and odd in the paramagnetic one.
 Note that the expressions (\ref{30}), (\ref{31}) for the tranfer
 matrix
 eigenvalues can be obtained from the representation
 (\ref{0}) for the partition function, if we expand the products (\ref{4a}):
 $$Z=Z^{(NS)}+Z^{(R)},$$
 $$Z^{(NS)}=\frac12(Q^{(NS,NS)}+Q^{(R,NS)})=\sum_{i}(\lambda_{i}^{NS})^{M},$$
$$Z^{(R)}=\frac12(Q^{(NS,R)}-Q^{(R,R)})=\sum_{i}(\lambda_{i}^{R})^{M}.$$

 Because of translation invariance, the transfer matrix can be diagonalized
 simultaneously with the translation operator. This additional requirement determines the orthonormal
 basis of eigenvectors up to permutations. It is convenient to interpret
 the elements of this basis in terms of NS-- and R--multiparticle states with
 appropriate
 values of quasimomenta: they are generated from the vacua
 $|\emptyset\rangle_{NS}$ and
 $|\emptyset\rangle_{R}$ by the action of fermionic creation operators.

 Let us consider for definiteness the ferromagnetic region and show
 how to express the two-point correlation function in terms of the matrix
 elements
 $$_{NS}\langle
 q_1,\ldots,q_{2K}|\hat{\sigma}(0,0)|p_1,\ldots,p_{2L}\rangle_R$$
 of Ising spin
 (the elements NS--NS and R--R vanish due to
 $\mathbb{Z}_2$-symmetry of the model):
 \begin{eqnarray}
 \nonumber\langle\sigma(0,0)\sigma(r_x,r_y)\rangle&=&
 _{NS}\langle\emptyset|\hat{\sigma}(0,0)\hat{\sigma}(r_x,r_y)|\emptyset\rangle_{NS}=\\
 \nonumber &=& _{NS}\langle\emptyset|\hat{\sigma}(0,0)\mathcal{T}^{r_x}
  \hat{\sigma}(0,r_y)\mathcal{T}^{-r_x}|\emptyset\rangle_{NS}=\\
  \nonumber &=& \sum\limits_{K=0}^{[N/2+1]}\frac{e^{-r_x/\Lambda}}{(2K)!}{\sum\limits_{p_1\ldots p_{2K}}}^{\!\!\!\!(R)}
  {\left|_{NS}\langle\emptyset|\hat{\sigma}(0,0)|p_1\ldots
  p_{2K}\rangle_R\right|}^{\;2}\times \\
  \label{32a}{} &\;&\qquad\; \times\;
  \exp\biggl(-r_x\sum\limits_{j=1}^{2K}\gamma(p_j)+ir_y\sum\limits_{j=1}^{2K}p_j\biggr).
 \end{eqnarray}
 Comparing this expression with the formulas (\ref{18})--(\ref{20}),
 we find spin matrix elements between the NS-vacuum and an arbitrary R--eigenstate:
 \be\label{32} {\left|_{NS}\langle\emptyset|\hat{\sigma}(0,0)|p_1\ldots
  p_{L}\rangle_R\right|}^{\;2}=\xi\xi_T
  \left[\frac{t_y\,(1-t_x^2)}{t_x(1-t_y^2)}\right]^{L^2/2} \prod\limits_{j=1}^{L}
  \frac{e^{-\eta(p_j)-\Lambda^{-1}}}{N\sinh\gamma(p_j)}\prod\limits_{i<j}^{L}
 \frac{\sin^{2}\frac{p_i-p_j}{2}}{\sinh^2\frac{\gamma(p_i)+\gamma(p_j)}{2}}\; .
 \ee
 The formula (\ref{32a}) then implies that  form factor expansion
 of the correlation function
 $\langle\sigma(0,0)\sigma(r_x,r_y)\rangle$ is obtained from
 (\ref{18})--(\ref{20}) by the substitution
 $$\frac{e^{-|r_x|\gamma(q_i)-\eta(q_i)}}{\sinh\gamma(q_i)}\rightarrow
 \frac{e^{-|r_x|\gamma(q_i)+ir_yq_i-\eta(q_i)}}{\sinh\gamma(q_i)}$$
 in the function $g_n$. An analogous result may be found in the paramagnetic case
 as well. Thus we have
 \begin{eqnarray}
 \label{33}\langle\sigma(0,0)\sigma(r_x,r_y)\rangle^{(-)}&=&\xi\xi_T\,
 e^{-|r_x|/\Lambda}\;\sum\limits_{l=0}^{[N/2]}g_{2l}(r_x,r_y),\\
 \label{34}\langle\sigma(0,0)\sigma(r_x,r_y)\rangle^{(+)}&=&\xi\xi_T\,
 e^{-|r_x|/\Lambda}
 \sum\limits_{l=0}^{[(N-1)/2]}g_{2l+1}(r_x,r_y),\end{eqnarray}
 where
 \be\label{35}g_n(r_x,r_y)=\frac{e^{-n/\Lambda}}{n!N^n}\;\left[
 \frac{t_y(1-t_x^2)}{t_x(1-t_y^2)}\right]^{n^2/2}{\sum\limits_q}^{(R)}\prod\limits_{j=1}^n
 \frac{e^{-|r_x|\gamma(q_j)+ir_yq_j-\eta(q_j)}}{\sinh\gamma(q_j)}\;\;F_{n}^{2}[q]\;.
 \ee

 As an illustration, consider the ``paramagnetic'' expansion for
 $K_y=0$. Notice that in this limit
 $$\cosh\gamma(q)=\coth2K_x,\;\;\;\gamma(q)=t_x=\mathrm{const},$$
 and, therefore
 $$\xi=\xi_T=1,\;\;\;\Lambda^{-1}=\eta(q)=0,$$
 $$\langle\sigma(0,0)\sigma(r_x,r_y)\rangle=\frac{1}{N}\;{\sum\limits_q}^{(b)}(t_x)^{|r_x|}
 e^{ir_yq}=(t_x)^{|r_x|}\;\delta_{\,0\,r_y}.$$
 As expected, the spins from different lattice rows are uncorrelated.

 Magnetic susceptibility of the two-dimensional Ising model on a finite $M\times
 N$ lattice in zero field may be written as a sum of correlation
 functions:
 $$\beta^{-1}\chi=\sum\limits_{r_x=0}^{M-1}\sum\limits_{r_y=0}^{N-1}
 \langle\sigma(0,0)\sigma(r_x,r_y)\rangle.$$
 In order to find the susceptibility on the cylinder, we can use the
 expression
 $$\beta^{-1}\chi=\sum\limits_{r_x=-\infty}^{\infty}\sum\limits_{r_y=0}^{N-1}
 \langle\sigma(0,0)\sigma(r_x,r_y)\rangle$$
 with some precautions \cite{we}. Because of  simple structure
 of the form factor expansions (\ref{33})--(\ref{35})
 this sum is easily computed, and we obtain the following expressions for the susceptibility
 in the ferromagnetic and paramagnetic region of parameters:
 $$\beta^{-1}\chi^{(-)}=\xi\xi_T \;\sum\limits_{l=0}^{[N/2]}\chi_{2l},\qquad
 \beta^{-1}\chi^{(+)}=\xi\xi_T
 \;\sum\limits_{l=0}^{[(N-1)/2]}\chi_{2l+1}\,,$$
 where
 $$\chi_n=\frac{e^{-n/\Lambda}}{n!N^{n-1}}\;\left[
 \frac{t_y(1-t_x^2)}{t_x(1-t_y^2)}\right]^{n^2/2}\times$$
 $$\times\;{\sum\limits_q}^{(R)}\prod\limits_{j=1}^n
 \frac{e^{-\eta(q_j)}}{\sinh\gamma(q_j)}\;F_{n}^{2}[q]\;\coth\biggl(\frac12
 \sum_{j=1}^n\gamma(q_j)+\Lambda^{-1}\biggr)\,
 \delta\biggl(\sum\limits_{j=1}^nq_j,\;0\;\;\mathrm{mod}\;\;2\pi\biggr), $$
 and $\delta$ is the Kronecker symbol.

 \section{Scaling limit}
 An important stage in the study of the two-dimensional Ising model is
 the analysis of its scaling limit \cite{palmer,mtw}. A new effect with respect to the
 case of the infinite plane is that the ``cylindrical parameters'' $\xi_{T}$, $\Lambda^{-1}$, $\eta$,
 which tend to zero for fixed distance from the critical point
 ($N\to\infty$, $\sinh2K-1=\mathrm{const}\neq0$), do not vanish
 in the scaling limit on the cylinder. Although the formulas for correlation
 functions obviously become  more involved,  it is possible to
 generalize several results obtained for the infinite plane to the
 case of the isotropic Ising model on the cylinder \cite{bugrij2}.
 For instance, it was shown \cite{I} that the Ising model
 correlation functions on the cylinder satisfy some integrable
 equations generalizing Painlev\'e~III and Painlev\'e~V equations
 obtained in the planar case. When considering the scaling
 limit of the anisotropic model on the cylinder, we do not
 encounter much difficulty provided we take into account some
 subtleties in the definition of scaling variables.

 Since in this case there are two parameters $K_{x}$, $K_{y}$,
 instead of a critical point we have the critical line $\sinh2K_{x}\sinh2K_{y}-1=0$.
 One has
 certain freedom in the definition of the distance from the
 critical line, which can be used to ensure that the results
 obtained in the anisotropic case in the scaling limit coincide
 with those obtained for the isotropic
 $(K_{x}=K_{y})$ model. It turns out that this condition is satisfied with the
 following choice of scaling variables:
 $$|r_x|\rightarrow\infty,\;\;\;|r_y|\rightarrow\infty,\;\;\;N\rightarrow\infty,\;\;\;
 \gamma(0)\rightarrow0,$$
 \be\label{36}\gamma(0)r_x=x=\mathrm{const},\;\;\;\gamma(0)r_y\sinh 2K_x=y=\mathrm{const},\;\;\;\gamma(0)N\sinh 2K_x=
 \beta=\mathrm{const}.\ee
 Because of the presence of the exponential factors $\e^{-|r_x|\gamma(q)+ir_yq}$ and  $\e^{-N\gamma(q)}$
 in the corresponding sums and integrals, we can suppose that
 $q\ll 1$ and, therefore,
 $$\gamma(q)=\gamma(0)\sqrt{1+\left(\frac{q}{\gamma(0)\sinh 2K_x}\right)^2}\;,\;\;\;
 \overline{\gamma}(q)=\gamma(0)\sinh2K_x\sqrt{1+\left(\frac{q}{\gamma(0)}\right)^2}\;.$$
 In order to find the scaling limit asymptotics of the cylindrical parameters
 $\Lambda$, $\xi_{T}$ and $\eta(q)$, it is convenient to use integral
 representations
 \begin{eqnarray*}
 \Lambda^{-1}&=&\frac{1}{\pi}\int\limits_0^\pi dp\;
 \ln\coth\frac{N\overline{\gamma}(p)}{2}\,,\\
 \eta(q)&=&\frac{1}{\pi}\int\limits_0^\pi dp\;
 \frac{\cos p-e^{-\gamma(q)}}{\cosh\gamma(q)-\cos
 p}\;\ln\coth\frac{N\overline{\gamma}(p)}{2}\,,\\
 \xi_T&=&\frac{N^2}{2\pi^2}\int\limits_0^\pi \frac{dp\;dq\;\overline{\gamma}\,'(p)\;\overline{\gamma}\,'(q)}
 {\sinh N\overline{\gamma}(p)\;\sinh N\overline{\gamma}(q)}\;\ln\left|\frac{\sin(p+q)/2}{\sin(p-q)/2}\right|,
 \end{eqnarray*}
 equivalent to the formulas (\ref{15}), (\ref{17}), (\ref{22}).
 Denoting
 $$\omega(q)=\sqrt{1+q^2}$$
 and using the above relations, we obtain the answer for the
 scaled
 correlation function of the 2D anisotropic Ising model on the
 cylinder:
  \begin{eqnarray}
  \label{37}\langle\sigma(0,0)\sigma(r_x,r_y)\rangle^{(-)}&=&\xi\tilde{\xi}_T(\beta)\,
 e^{-|x|/\tilde{\Lambda}(\beta)}\;\sum\limits_{n=0}^{\infty}\tilde{g}_{2n}(x,y,\beta)
 \;,\\
 \label{38}\langle\sigma(0,0)\sigma(r_x,r_y)\rangle^{(+)}&=&\xi\tilde{\xi}_T(\beta)\,
 e^{-|x|/\tilde{\Lambda}(\beta)}\;\sum\limits_{n=0}^{\infty}\tilde{g}_{2n+1}(x,y,\beta)
 \;,\end{eqnarray}
 \be\label{39}\tilde{g}_n(x,y,\beta)=\frac{1}{n!\beta^n}\;{\sum\limits_{[l]}}\prod\limits_{j=1}^n
 \frac{e^{-|x|\omega(q_j)+iyq_j-\tilde{\eta}(q_j,\beta)}}{\omega(q_j)}\;\;\tilde{F}_{n}^{2}[l]\;.
 \ee
 Here, the summation is performed over integer $l$
 $$\sum\limits_{[l]}=\sum\limits_{l_1=-\infty}^{\infty}\ldots\sum\limits_{l_n=-\infty}^{\infty},$$
 the quasimomenta take on the bosonic values $q_j=\frac{2\pi
 l_j}{\beta}$, and the cylindrical parameters and form factors are given by
 \begin{eqnarray*}
 \tilde{\Lambda}^{-1}(\beta)&=&\frac{1}{\pi}\int\limits_0^\infty dp\;
 \ln\coth\frac{\beta\omega(p)}{2}\,,\\
 \tilde{\eta}(q_j,\beta)&=&\frac{1}{\pi}\int\limits_0^\infty dp\;
 \frac{2\omega(q_j)}{\omega^2(q_j)+p^2}\;\ln\coth\frac{\beta\omega(p)}{2}\,,\\
 \tilde{\xi}_T(\beta)&=&\frac{\beta^2}{2\pi^2}\int\limits_0^\infty
 \frac{dp\;dq\;\omega\,'(p)\;\omega\,'(q)}
 {\sinh \beta\,\omega(p)\;\sinh \beta\omega(q)}\;\ln\left|\frac{p+q}{p-q}\right|,
 \end{eqnarray*}
 $$\tilde{F}_n[l]=\prod\limits_{1\leq i<j\leq n}\frac{q_i-q_j}{\omega(q_i)+\omega(q_j)}\;.$$
 We see that the spin-spin correlation function on the cylinder in the scaling limit
 defined by (\ref{36}) is given by the same functions of renormalized coordinates as in
 the isotropic model. Therefore, all results of \cite{I} (determinant representations of
 the correlation fuctions, differential equations) apply to the anisotropic case as well.

 \section{Spin matrix elements}
 In order to calculate multipoint correlation functions on the cylinder
 and torus, it is necessary (and sufficient) to have formulas not only for
 the form factors
 ${\left|_{NS}\langle\emptyset|\hat{\sigma}(0,0)|p_1\ldots
  p_{L}\rangle_R\right|}$, but also for
 all other spin matrix elements. For instance, the two-point correlation function
 in the Ising model on the periodic lattice of size $M\times N$ in the ferromagnetic region
 of parameters can be written as follows:
$$\langle\sigma(0,0)\sigma(r_x,r_y)\rangle=\frac{
 \mathrm{Tr}\left\{\hat{\sigma}(0,0)
 \mathcal{T}^{r_x}
  \hat{\sigma}(0,r_y)\mathcal{T}^{M-r_x}\right\}}{\mathrm{Tr}\;\mathcal{T}^M}=$$
  $$
 =\sum\limits_{K=0}^{[N/2+1]}\sum\limits_{L=0}^{[N/2+1]}
 {\sum\limits_{q_1\ldots q_{2K}}}^{\!\!\!\!(NS)}
 {\sum\limits_{p_1\ldots p_{2L}}}^{\!\!\!\!(R)}
  {\left|_{NS}\langle q_1\ldots q_{2K}|\hat{\sigma}(0,0)|p_1\ldots
  p_{2L}\rangle_R\right|}^{\;2}\;\frac{e^{ir_y\sum\limits_{j=1}^{2K}q_j-ir_y\sum\limits_{j=1}^{2L}p_j}}{(2K)!(2L)!}\times$$
  $$\times  \biggl\{
  e^{-(M-r_x)\sum\limits_{j=1}^{2K}\gamma(q_j)-r_x\bigl(\Lambda^{-1}+\sum\limits_{j=1}^{2L}\gamma(p_j)\bigr)}+
  e^{-r_x\sum\limits_{j=1}^{2K}\gamma(q_j)-(M-r_x)\bigl(\Lambda^{-1}+\sum\limits_{j=1}^{2L}\gamma(p_j)\bigr)}
  \biggr\}\biggm/\mathrm{Tr}\,\left(\frac{\mathcal{T}}{\lambda_0}\right)^M,$$
  where $\lambda_0$ denotes the largest transfer matrix eigenvalue, which
  corresponds to the eigenvector
  $|\emptyset\rangle_{NS}$.

  In \cite{angers}, we have proposed a formula for arbitrary spin matrix elements of the isotropic Ising model
  on a finite periodic lattice. This formula can be easily generalized to the anisotropic
  case, using two hints from the above:
  \begin{itemize}
  \item assume that all the difference due to anisotropy consists
  in the
  redefinition of the function $\gamma(q)$ in the formula for spin matrix element,
  and in the appearance of the factor $\displaystyle\frac{t_y(1-t_x^2)}{t_x(1-t_y^2)}$ to some power
   (see, e.~g., the formula (\ref{32}));
  \item also assume that in the scaling limit (\ref{36})
  \textit{all} multipoint correlation functions are given by the
  same formulas
  \cite{bugrij2,angers}
  as in the isotropic case.
  \end{itemize}
  Omitting the calculations, we present only the final expression for the spin
  matrix element, which follows from the above assumptions:
   $$ {\left|_{NS}\langle q_1\ldots q_K|\hat{\sigma}(0,0)|p_1\ldots
  p_{L}\rangle_R\right|}^{\;2}=\xi\xi_T
  \prod\limits_{j=1}^{K}
  \frac{e^{\eta(q_j)+\Lambda^{-1}}}{N\sinh\gamma(q_j)}
  \prod\limits_{j=1}^{L}
  \frac{e^{-\eta(p_j)-\Lambda^{-1}}}{N\sinh\gamma(p_j)}\times$$
  \be\label{40}\times
  \left[\frac{t_y\,(1-t_x^2)}{t_x(1-t_y^2)}\right]^{\frac{(K-L)^2}{2}}
  \prod\limits_{i<j}^{K}
  \frac{\sin^{2}\frac{q_i-q_j}{2}}{\sinh^2\frac{\gamma(q_i)+\gamma(q_j)}{2}}
  \prod\limits_{i<j}^{L}
 \frac{\sin^{2}\frac{p_i-p_j}{2}}{\sinh^2\frac{\gamma(p_i)+\gamma(p_j)}{2}}
  \prod\limits_{\substack{1\leq i\leq K \\ 1\leq j\leq L}}
 \frac{\sinh^2 \frac{\gamma(q_i)+\gamma(p_j)}{2}}{\sin^2
 \frac{q_i-p_j}{2}}\, . \ee
 As in the isotropic model case, this representation was verified for
 finite-row Ising chains with  $N=1,2,3,4$.
 The scaling limit of the expression (\ref{40}) obviously coincides with
 the classical result \cite{BK} in the limit of continuous infinite plane, and with the
 corresponding results on the continuous infinite cylinder
 \cite{fonseca}.

 \section{Discussion}
 Lattice systems with cylindrical geometry recently became important for nanoelectronics.
 However, the elementary cell in these systems is hexagonal. The star-triangle transformation
 brings the Ising model defined on such a lattice to the Ising model on a triangular
 lattice. The simple form of the expressions obtained in the present paper gives a hope
 that similar results can also be obtained in the latter case.

 Exact expressions for the correlation functions are also known for several 1D quantum systems,
 in particular, for the XXZ Heisenberg spin chain \cite{maillet}. In these models, matrix elements
 of local operators on the hamiltonian eigenstates were calculated using the Bethe
 ansatz. One can try to use our results to analyze the behaviour
 of such integrable one-dimensional models in a finite volume
 and/or at non-zero temperature.
 \vspace{0.5cm}

 We thank S. Z. Pakuliak, V. N. Roubtsov and V. N. Shadura
 for their interest to this work and useful discussions. This work was supported
 by the INTAS program under grant No. 03513350.

 \bibliographystyle{plain}

\end{document}